\documentclass[ALICE,manyauthors]{cernphprep}

\usepackage{graphicx, subfigure}
\usepackage{amssymb}
\usepackage{amsmath}
\usepackage{lineno}
\usepackage{cite}
\usepackage{mciteplus}
\usepackage{hyperref}

\usepackage{multirow}

\begin{document}%

%
\begin{titlepage}
\PHnumber{2012-306}               
\PHdate{Oct 16, 2012}              

\title{Transverse momentum distribution and nuclear modification factor 
of charged particles in p--Pb collisions at  
$\sqrt{s_{\rm{NN}}} = 5.02$ TeV}

\ShortTitle{Charged-particle transverse momentum distribution in p--Pb}   

\Collaboration{ALICE Collaboration%
         \thanks{See Appendix~\ref{app:collab} for the list of collaboration
                      members}}
\ShortAuthor{ALICE Collaboration} 

\begin{abstract}
The transverse momentum ($p_{\mathrm T}$) distribution of primary 
charged particles is measured in minimum bias (non-single-diffractive)
\mbox{p--Pb} collisions at \mbox{$\sqrt{s_{\mathrm{NN}}}=5.02$~TeV} 
with the ALICE detector at the LHC.  
The $p_{\mathrm T}$ spectra measured near central rapidity in the range 
\mbox{$0.5<p_{\mathrm T}<20$~GeV/$c$} 
exhibit a weak pseudorapidity dependence.
The  nuclear modification factor $R_{\mathrm{pPb}}$ 
is consistent with unity for $p_{\mathrm T}$ above 2~GeV/$c$.
This measurement indicates that the strong suppression of hadron 
production at high $p_{\mathrm T}$ observed in Pb--Pb collisions at the LHC 
is not due to an initial-state effect.
The measurement is compared to theoretical calculations.
\end{abstract}
\end{titlepage}
\setcounter{page}{2}
Measurements of particle production in proton-nucleus collisions 
at high energies allow the study of fundamental properties of Quantum 
Chromodynamics (QCD) at low parton fractional momentum $x$ and
high gluon densities
(see \cite{Salgado:2011wc} for a recent review). 
They also provide a reference measurement for the studies of deconfined matter 
created in nucleus-nucleus collisions \cite{Muller:2012zq}.

Parton energy loss in hot QCD matter is expected to lead to a
modification of energetic jets in this medium (jet quenching)
\cite{Bjorken:1982tu}.
Originating from energetic partons produced in initial hard collisions, 
hadrons at high transverse momentum $p_{\mathrm T}$ are an important observable
for the study of deconfined matter.
Experiments at RHIC have shown \cite{Adcox:2001jp,Adler:2002xw} 
that the production of charged hadrons at high $p_{\mathrm T}$
in Au--Au collisions is suppressed compared to the expectation from 
an independent superposition of nucleon--nucleon collisions (binary collision 
scaling). 

By colliding Pb nuclei at the LHC it was shown 
\cite{Aamodt:2010jd,CMS:2012aa,Abelev:2012eq}
that the production of charged hadrons in central collisions at a 
center-of-mass (cms) collision energy per nucleon pair
$\sqrt{s_{\mathrm{NN}}} = 2.76$ TeV shows a stronger suppression than at RHIC, 
indicating a state of QCD matter with an even higher energy density.
At the LHC, the suppression remains substantial up to 100 GeV/$c$ 
\cite{CMS:2012aa,Abelev:2012eq} and is also seen in reconstructed jets 
\cite{Aad:2012is}.
A p--Pb control experiment is needed to establish whether the initial 
state of the colliding nuclei plays a role in the observed suppression 
of hadron production at high-$p_{\mathrm T}$ in Pb--Pb collisions.
In addition, p--Pb data should also provide tests of models that 
describe QCD matter at high gluon density, giving 
insight into phenomena such as parton shadowing or gluon saturation 
\cite{Salgado:2011wc}. 

In this letter, we present a measurement of the  $p_{\mathrm T}$ distributions of
charged particles in \mbox{p--Pb} collisions at $\sqrt{s_{\rm{NN}}}=5.02$~TeV.
The data were recorded with the ALICE detector \cite{Aamodt:2008zz}  
during a short LHC p--Pb run performed in September 2012
in preparation for the main run scheduled at the beginning of 2013. 
Each beam contained 13 bunches; 8 pairs of bunches were colliding in 
the ALICE interaction region, providing a luminosity of about 
$8\times10^{25}$ cm$^{-2}$s$^{-1}$.
The interaction region had an r.m.s. width of 6.3 cm in the longitudinal 
direction and of about 60~$\mu$m in the transverse directions.

The trigger required a signal in either of two arrays of 32 scintillator tiles 
each, covering full azimuth and $2.8<\eta_{\mathrm{lab}}<5.1$~(VZERO-A) and 
$-3.7<\eta_{\mathrm{lab}}<-1.7$~(VZERO-C), respectively.
The pseudorapidity in the detector reference frame,
\begin{math} 
\eta_{\mathrm{lab}} = -{\rm{ln}}[{\rm{tan}} (\theta/2)]
\end{math}, 
with $\theta$ the polar angle between the charged particle
and the beam axis, is defined such that the proton beam has negative 
$\eta_{\mathrm{lab}}$.
This configuration led to a trigger rate of about 200~Hz, 
with a hadronic collision rate of about 150~Hz.
The efficiency of the VZERO trigger was estimated from a control sample 
of events triggered by signals from two Zero Degree Calorimeters~(ZDC)
positioned symmetrically at 112.5~m from the interaction point, 
with an energy resolution of about 20\% for single neutrons of a few TeV.

The offline event selection is identical to that used for the analysis of 
charged-particle pseudorapidity density 
($\mathrm{d} N_{\mathrm{ch}}/\mathrm{d} \eta_{\mathrm{lab}}$) 
reported in \cite{alice:2012ppb}.
A signal is required in both VZERO-A and VZERO-C.
Beam--gas and other machine-induced background events with deposited energy 
above the thresholds in the VZERO or ZDC detectors are suppressed by 
requiring the signal timing to be compatible with that of a nominal p--Pb 
interaction.
The remaining background after these requirements is estimated from 
triggers on non-colliding bunches, and found to be negligible.
The resulting sample of events consists of non-single-diffractive (NSD) 
collisions as well as single-diffractive and electromagnetic interactions.
The efficiency of the trigger and offline event selection for the different 
interactions is estimated using a combination of event generators, see 
\cite{alice:2012ppb} for details.
An efficiency of 99.2\% for NSD collisions is estimated, with 
a negligible contamination from single-diffractive and electromagnetic 
interactions. The number of events used for the analysis is 
$1.7\times10^{6}$.

The primary vertex position is determined with tracks reconstructed in 
the Inner Tracking System (ITS) and the Time Projection Chamber (TPC) by 
using a $\chi^2$ minimization procedure described in \cite{Abelev:2012eq}. 
The event vertex reconstruction algorithm is fully efficient for 
events with at least one track in the acceptance, $|\eta_{\mathrm{lab}}| < 1.4$  
(when the center of the interaction region is included as an additional 
constraint).
An event is accepted if the coordinate of the reconstructed vertex measured 
along the beam direction is within \mbox{$\pm10$~cm} around the center of 
the interaction region.

Primary charged particles are defined as all prompt particles produced
in the collision, including decay products, except those from weak
decays of strange hadrons. Selections based on the number of space points 
and the quality of the track fit, as well as on the distance of closest 
approach to the reconstructed vertex, are applied to the reconstructed tracks 
(see \cite{Abelev:2012eq} for details).
The efficiency and purity of the primary charged particle selection are 
estimated from a Monte Carlo simulation using the DPMJET event generator 
\cite{Roesler:2000he}
with particle transport through the detector using GEANT3~\cite{GEANT3}. 
The systematic uncertainties on corrections are estimated via a comparison
to a Monte Carlo simulation using the HIJING event generator 
\cite{Wang:1991hta}.
The overall primary charged particle reconstruction efficiency (the product
of tracking efficiency and acceptance) for $|\eta_{\mathrm{lab}}|<0.8$ is 
79\% at $p_{\mathrm T} =0.5$~GeV/$c$, reaches 81\% at 0.8 GeV/$c$ and decreases 
to 72\% for $p_{\rm T}>2$~GeV/$c$.
From Monte Carlo simulations it is estimated that the residual contamination 
from secondary particles is 1.6\% at $p_{\mathrm T} =0.5$~GeV/$c$
and decreases to about 0.6\% for $p_{\mathrm T}>2$~GeV/$c$. 

The transverse momentum of charged particles is determined from the track
curvature  in the magnetic field of $0.5$~T.  
The $p_{\mathrm T}$ resolution is estimated from the space-point 
residuals to the track fit and verified by the width of the invariant mass of 
K$^0_{\mathrm{S}}$ mesons reconstructed in their decay to two charged pions. 
For the selected tracks the relative $p_{\mathrm T}$ resolution  
is 1.3\% at  $p_{\mathrm T}=0.5$~GeV/$c$, 
has a minimum of 1.0\% at $p_{\mathrm T}=1$~GeV/$c$, and increases linearly 
to 2.2\% at $p_{\mathrm T}=20$~GeV/$c$. 
The uncertainty on the $p_{\mathrm T}$ resolution is $\pm 0.7$\% at 
$p_{\mathrm T}=20$~GeV/$c$, leading to a systematic uncertainty on the 
differential yield of up to 3\% at this $p_{\mathrm T}$ value.

\begin{table}[htb]
\centering
\caption{Systematic uncertainties on the $p_{\mathrm T}$-differential yields 
in p--Pb and pp collisions for $|\eta_{\mathrm{cms}}|<0.3$. The quoted ranges
span the $p_{\mathrm T}$ dependence of the uncertainties.}
\label{tab_systerr} 
\begin{tabular}{ll}
Uncertainty & Value \\ \hline
Event selection & 1.0--2.0\% \\
Track selection & 0.9--2.7\% \\
Tracking efficiency & 3.0\% \\
$p_{\mathrm T}$ resolution & 0--3.0\%  \\
Particle composition & 2.2--3.1\% \\
MC generator used for correction & 1.0\% \\
Secondary particle rejection & 0.4--1.1\% \\ 
Material budget & 0--0.5\% \\ 
Acceptance (conversion to $\eta_{\mathrm{cms}}$) & 0--0.6\% \\

\hline
Total for p--Pb, $p_{\mathrm T}$-dependent & 5.2--5.5\% \\
\hline
Normalization p--Pb & 3.1\% \\
\hline
\hline
Total for pp, $p_{\mathrm T}$-dependent  & 7.7--8.2\% \\
Normalization pp & 3.6\% \\
Nuclear overlap $\langle T_{\mathrm{pPb}} \rangle$  & 3.6\% \\

\end{tabular}
\end{table}

Due to the different energy per nucleon of the two colliding beams,
imposed by the two-in-one magnet design of the LHC,
the nucleon-nucleon cms moves with a rapidity 
$y_{\mathrm{NN}}=0.465$ in the direction of the proton beam.
As a consequence, the detector coverage, $|\eta_{\mathrm{lab}}| < 0.8$, implies, 
for the nucleon-nucleon cms, roughly \mbox{$-0.3<\eta_{\mathrm{cms}}<1.3$}.
The calculation of $\eta_{\mathrm{cms}}=\eta_{\mathrm{lab}} + y_{\mathrm{NN}}$ 
is accurate only for massless particles or at high $p_{\mathrm T}$.
Consequently, the differential yield at low $p_{\mathrm T}$ suffers from 
a distortion, which is estimated and corrected for based on the particle 
composition in the HIJING event generator.
For $p_{\mathrm T}=0.5$ GeV/$c$, the correction is 1\% for 
$|\eta_{\mathrm{cms}}|<0.3$  and reaches 3\% for $0.8<\eta_{\mathrm{cms}}<1.3$.
The systematic uncertainties were estimated by varying the 
relative particle abundances by factors of 2 around the nominal values.
The uncertainty is sizable only at low $p_{\mathrm T}$ and is dependent
on $\eta_{\mathrm{cms}}$. It is 0.6\% for $|\eta_{\mathrm{cms}}|<0.3$,  
4.3\% for $0.3<\eta_{\mathrm{cms}}<0.8$, 
and 5.1\% for $0.8<\eta_{\mathrm{cms}}<1.3$.

The systematic uncertainties on the $p_{\mathrm T}$ spectrum are summarized in 
Table~\ref{tab_systerr} for $|\eta_{\mathrm{cms}}|<0.3$. 
The total uncertainties exhibit a weak $p_{\mathrm T}$ and $\eta_{\mathrm{cms}}$ 
dependence.
The total systematic uncertainties range between 5.2\% and 5.5\% for 
$|\eta_{\mathrm{cms}}|<0.3$ and reach between 5.6\% and 7.1\% for 
$0.8<\eta_{\mathrm{cms}}<1.3$.

In order to quantify nuclear effects in p--Pb collisions, a comparison
to a reference $p_{\mathrm T}$ spectrum in pp collisions is needed.
In the absence of a measurement at $\sqrt{s}=$5.02 TeV, the reference spectrum 
is obtained by interpolating or scaling data measured at $\sqrt{s}=$2.76 
and 7 TeV.
For $p_{\mathrm{T}} < 5$ GeV/$c$, the measured invariant cross section for 
charged particle production in inelastic pp collisions, 
${\mathrm d}^{2}\sigma^{\mathrm{pp}}_{\mathrm{ch}}/{\mathrm d}\eta {\mathrm d}p_{\mathrm T}$,
is interpolated bin-by-bin, assuming a power law dependence as a function 
of $\sqrt{s}$. 
For $p_{\mathrm{T}} > 5$ GeV/$c$, the measured data at $\sqrt{s} = 7$ TeV is 
scaled by a factor obtained from next-to-leading order (NLO) perturbative QCD 
calculations \cite{Sassot:2010bh}.
For $p_{\mathrm{T}} < 5$ GeV/$c$, the largest of the relative systematic 
uncertainties of the spectrum at 2.76 or 7 TeV is assigned as the 
systematic uncertainty at the interpolated energy. 
For $p_{\mathrm{T}} > 5$ GeV/$c$, the relative difference between 
the NLO-scaled spectrum for different choices of the renormalization $\mu_R$
and factorization $\mu_F$ scales ($\mu_R=\mu_F=p_{\mathrm{T}}$, 
$p_{\mathrm{T}}/2$, $2p_{\mathrm{T}}$)
is added to the systematic uncertainties on the spectrum at 7 TeV. 
In addition, an uncertainty of 2.2\% is estimated comparing the interpolated 
and the NLO-scaled data. 
The total systematic uncertainty range from 7.7\% to 8.2\% for 
$0.5<p_{\mathrm{T}}<20$ GeV/$c$. 
The NLO-based scaling of the data at $\sqrt{s} = 2.76$ TeV gives a result well
within these uncertainties.
More details can be found in \cite{pp_ref:2012aa}. 

\begin{figure}[htb]
\begin{center}
\includegraphics[width=0.52\textwidth]{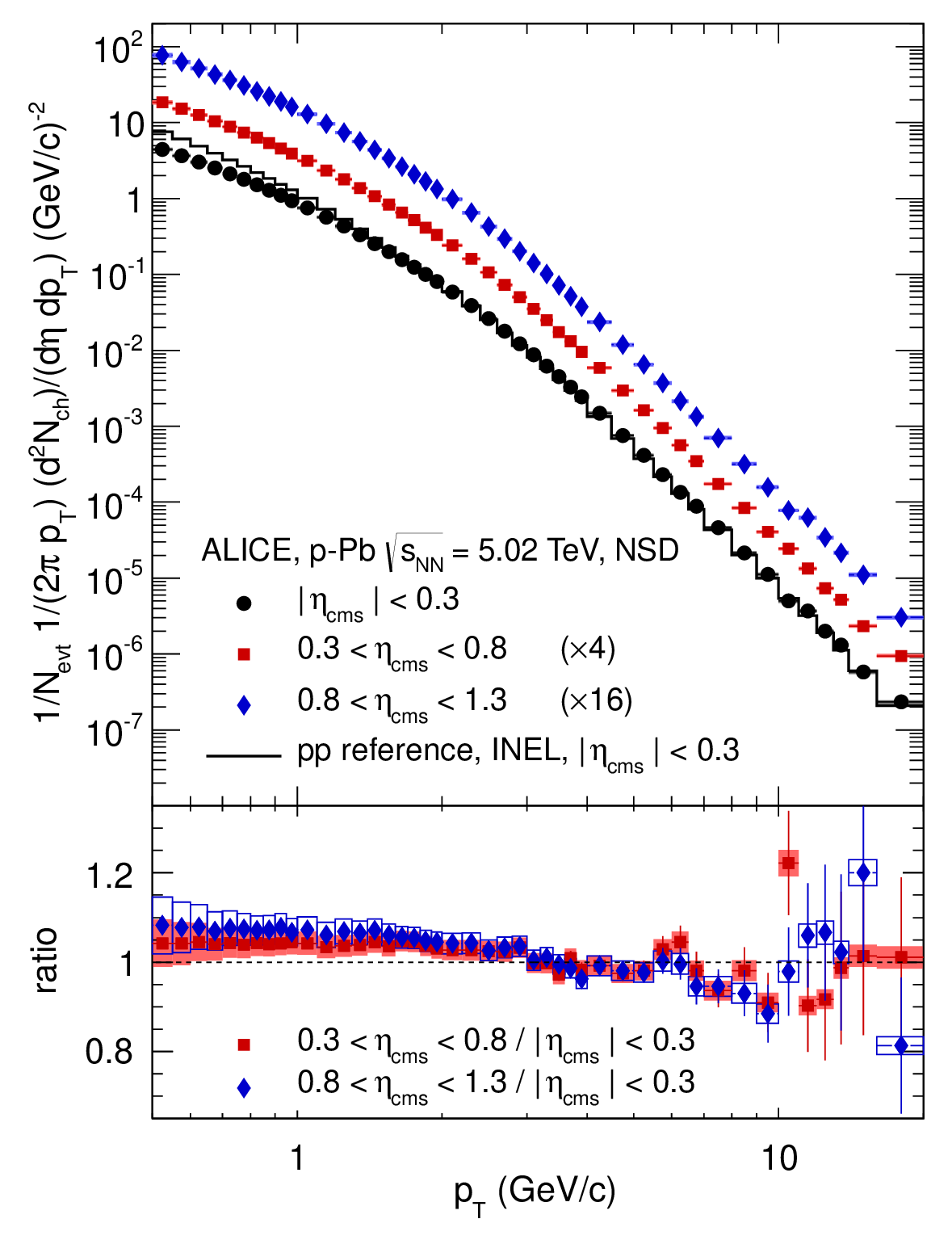}
\end{center}
\caption{Transverse momentum distributions of charged particles 
 in minimum bias (NSD) \mbox{p--Pb} collisions for different pseudorapidity 
ranges (upper panel).
The spectra are  scaled  by the factors indicated.  
The histogram represents the reference spectrum in inelastic (INEL) pp 
collisions (see text).
The lower panel shows the ratio of the spectra at forward pseudorapidities
to that at $|\eta_{\mathrm{cms}}|<0.3$. 
The vertical bars (boxes) represent the statistical (systematic) errors.
}
\label{fig_pt} 
\end{figure}

The final pp reference spectrum is the product of the interpolated invariant 
cross section and the average nuclear overlap $\langle T_{\mathrm{pPb}}\rangle$,
calculated employing the Glauber model \cite{Alver:2008aq}, which gives 
$\langle T_{\mathrm{pPb}}\rangle=\langle N_{\mathrm{coll}}\rangle / \sigma_{\mathrm{NN}}
=0.0983 \pm 0.0035$ mb$^{-1}$, with $\langle N_{\mathrm{coll}}\rangle=6.9\pm 0.7$ 
and $\sigma_{\mathrm{NN}}=70\pm 5$ mb.
The uncertainty is obtained by varying the parameters in the Glauber model 
calculation, see \cite{alice:2012ppb} (the uncertainties on $\sigma_{\mathrm{NN}}$
and $\langle N_{\mathrm{coll}}\rangle$ cancel partially in the calculation of 
$\langle T_{\mathrm{pPb}}\rangle$).

The  $p_{\mathrm T}$ spectra of charged particles measured in minimum bias 
(0-100\% centrality, NSD) 
\mbox{p--Pb} collisions at \mbox{$\sqrt{s_{\rm{NN}}} = 5.02$~TeV} are shown in 
Fig.~\ref{fig_pt} together with the interpolated pp reference spectrum.
At high $p_{\mathrm T}$, the $p_{\mathrm T}$ distributions in \mbox{p--Pb} 
collisions are similar to that in pp collisions, as expected in the 
absence of nuclear effects. 
There is an indication of a softening of the $p_{\mathrm T}$ spectrum when 
going from central to forward pseudorapidity.
This is a small effect, as seen in the ratios of the spectra for forward 
pseudorapidities to that at  $|\eta_{\mathrm{cms}}|<0.3$, shown in 
Fig.~\ref{fig_pt} (lower panel). 
We note that, that several contributions to the systematic uncertainties 
cancel in the ratios, resulting in systematic uncertainties of 2.2-5.2\% 
(2.2-5.9\%) for the ratio of the spectrum in $0.3<\eta_{\mathrm{cms}}<0.8$
($0.8<\eta_{\mathrm{cms}}<1.3$) to that in $|\eta_{\mathrm{cms}}|<0.3$.
Calculations with the DPMJET event generator \cite{Roesler:2000he}, which
predict well the measured $\mathrm{d} N_{\mathrm{ch}}/\mathrm{d} \eta_{\mathrm{lab}}$ 
\cite{alice:2012ppb}, overpredict the spectra by up to 22\% for 
$p_{\mathrm{T}}<0.7$ GeV/$c$ and underpredict them by up to 50\% for 
$p_{\mathrm{T}}>0.7$ GeV/$c$.

In order to quantify nuclear effects in p--Pb collisions, the 
$p_{\mathrm T}$-differential yield relative to the pp reference, the nuclear 
modification factor, is calculated as:
\begin{equation}\label{eq_RpPb}
R_{\mathrm{pPb}} (p_{\mathrm T} ) = \frac{{\mathrm d}^{2}N^{\mathrm{pPb}}_{\mathrm{ch}}/
{\mathrm d}\eta{\mathrm d}p_{\mathrm{T}} }{\langle T_{\mathrm{pPb}} \rangle 
{\mathrm d}^{2}\sigma^{\mathrm{pp}}_{\mathrm{ch}}/{\mathrm d}\eta {\mathrm d}p_{\mathrm T} },
\end{equation}
where $N^{\mathrm{pPb}}_{\rm{ch}}$ is the charged particle yield in p--Pb 
collisions. 
The nuclear modification factor is unity for hard processes
which are expected to exhibit binary collision scaling.
For the region of several tens of GeV, binary collision scaling was  
experimentally confirmed in Pb--Pb collisions at the LHC by the recent 
measurements of observables which are not affected by hot QCD matter, 
direct photon \cite{Chatrchyan:2012vq}, Z$^0$ \cite{Chatrchyan:2011ua}, 
and W$^{\pm}$ \cite{Chatrchyan:2012nt} production.
The present measurement in p--Pb collisions extends this important 
experimental verification down to the GeV scale and to hadronic observables.

\begin{figure}[htb]
\begin{center}
\includegraphics[width=0.55\textwidth]{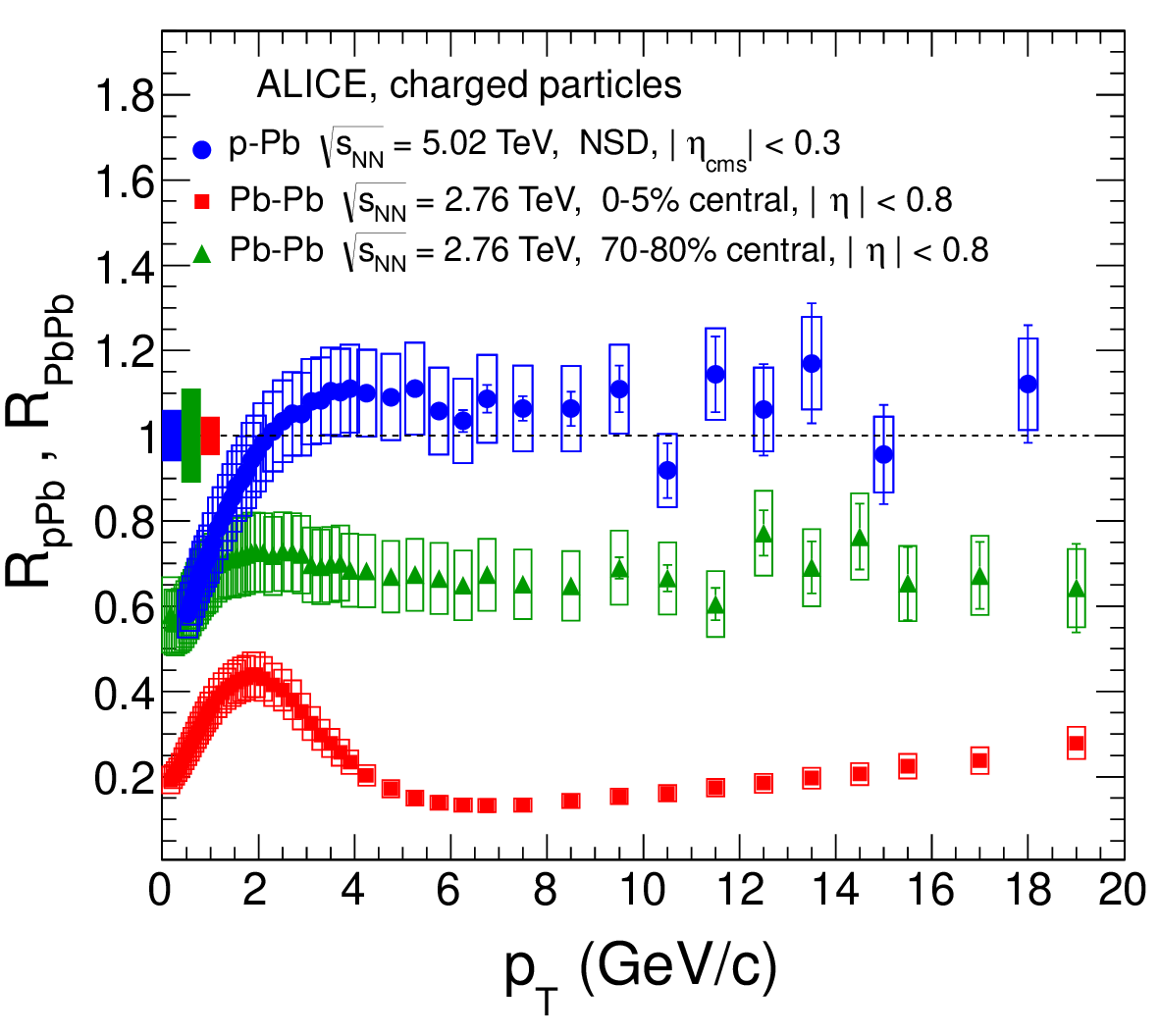}
\end{center}
\caption{The nuclear modification factor of charged particles as a function 
of transverse momentum in minimum bias (NSD) \mbox{p--Pb} collisions at 
$\sqrt{s_{\mathrm{NN}}} = 5.02$~TeV. 
The data for $|\eta_{\mathrm{cms}}|<0.3$ are compared to measurements 
\cite{Abelev:2012eq} in central (0--5\% centrality) and peripheral (70--80\%) 
Pb--Pb collisions at $\sqrt{s_{\mathrm{NN}}} = 2.76$~TeV.
The statistical errors are represented by vertical bars, the systematic 
errors by (filled) boxes around data points.
The relative systematic uncertainties on the normalization are shown as boxes 
around unity  near $p_{\mathrm{T}}=0$ for p--Pb (left box), peripheral Pb--Pb
 (middle box) and central Pb-Pb (right box).
}
\label{fig_raa_Pb} 
\end{figure}

The measurement of the nuclear modification factor $R_{\mathrm{pPb}}$ for 
charged particles at \mbox{$|\eta_{\mathrm{cms}}|<0.3$}, is shown in 
Fig.~\ref{fig_raa_Pb}. 
The uncertainties of the p--Pb and pp spectra are added in quadrature,
separately for the statistical and systematic uncertainties.
The total systematic uncertainty on the normalization, quadratic sum of 
the uncertainty on $\langle T_{\mathrm{pPb}} \rangle$, the normalization of the 
pp data and the normalization of the p--Pb data, amounts to 6.0\%.

In Fig.~\ref{fig_raa_Pb} we compare the measurement of the nuclear
modification factor in p--Pb to that in central (0--5\% centrality)
and peripheral (70--80\% centrality) Pb--Pb collisions at 
$\sqrt{s_{\mathrm{NN}}} = 2.76$~TeV \cite{Abelev:2012eq}.   
$R_{\mathrm{pPb}}$ is consistent with unity for $p_{\mathrm{T}}\gtrsim 2$ GeV/$c$,
demonstrating that the strong suppression observed in central Pb--Pb collisions 
at the LHC \cite{Aamodt:2010jd,CMS:2012aa,Abelev:2012eq} 
is not due to an initial-state effect, but rather a fingerprint of the hot 
matter created in collisions of heavy ions. 

The so-called Cronin effect \cite{PhysRevD.11.3105} (see \cite{Accardi:2002ik}
for a review), namely a nuclear modification factor above unity at intermediate
$p_{\mathrm{T}}$, was observed at lower energies in proton--nucleus collisions.
In d--Au collisions at \mbox{$\sqrt{s_{\rm{NN}}}=200$~GeV}, $R_{\mathrm{dAu}}$ 
reached values of about 1.4 for charged hadrons in the $p_{\mathrm{T}}$ range
3 to 5 GeV/$c$ \cite{Adler:2003ii,Adams:2003im,Arsene:2003yk,Back:2004bq}.
The present measurement clearly indicates a smaller magnitude of the Cronin 
effect at the LHC; the data are even consistent with no enhancement
within systematic uncertainties.

\begin{figure}[t]
\begin{center}
\includegraphics[width=0.52\textwidth]{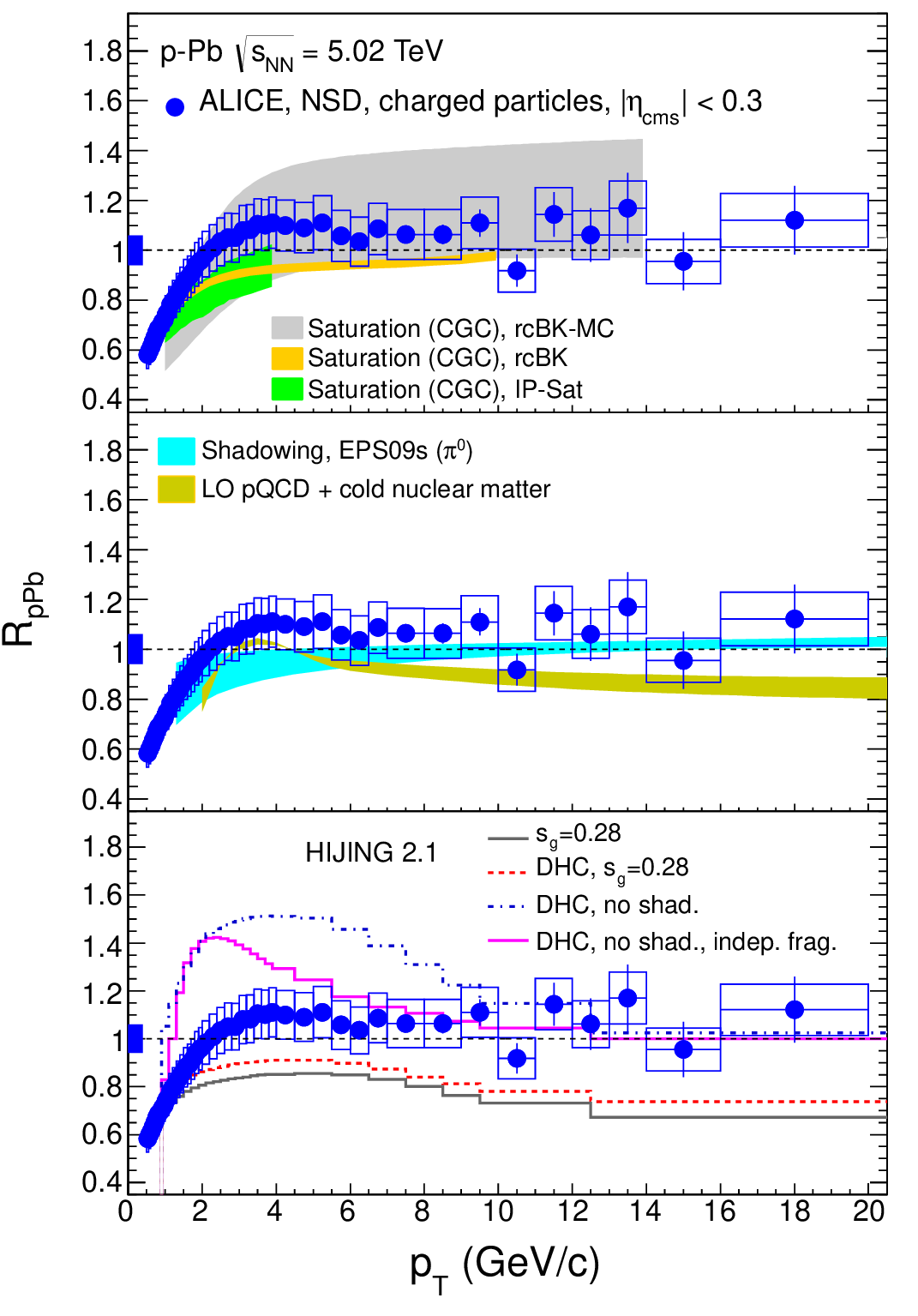}
\end{center}
\caption{Transverse momentum dependence of the nuclear modification 
factor $R_{\mathrm{pPb}}$ of charged particles measured in minimum bias (NSD) 
\mbox{p--Pb} collisions at $\sqrt{s_{\mathrm{NN}}} = 5.02$~TeV. 
The ALICE data in $|\eta_{\mathrm{cms}}|<0.3$ (symbols) are compared to model 
calculations (bands or lines, see text for details).
The vertical bars (boxes) show the statistical (systematic) errors.
The relative systematic uncertainty on the normalization is shown as a box 
around unity near $p_{\mathrm{T}}=0$. 
}
\label{fig_raa} 
\end{figure}

Data in p--Pb are important also to provide constraints to models.
For illustration, in Fig.~\ref{fig_raa} the measurement of $R_{\mathrm{pPb}}$ 
at \mbox{$|\eta_{\mathrm{cms}}|<0.3$} is compared to theoretical predictions.
Note that the measurement is performed for NSD collisions.
With the HIJING \cite{Wang:1991hta} and DPMJET \cite{Roesler:2000he} event 
generators, it is estimated that the inclusion of single-diffractive events
would lead to a decrease of $R_{\mathrm{pPb}}$ by 3--4\%.
Several predictions based on the saturation (Colour Glass Condensate, CGC) 
model are available \cite{Tribedy:2011aa,Albacete:2012xq,Rezaeian:2012ye}.
The calculations of Tribedy and Venugopalan \cite{Tribedy:2011aa}
are shown for two implementations (rcBK and IP-Sat, see \cite{Tribedy:2011aa} 
for details). The calculations within IP-Sat are consistent with the data, 
while those within rcBK slightly underpredict the measurement.
The prediction of Albacete et al. \cite{Albacete:2012xq}, for the rcBK Monte
Carlo model (rcBK-MC), is consistent with the measurement within the rather 
large uncertainties of the model. 
The CGC calculations  of Rezaeian \cite{Rezaeian:2012ye}, not included in 
Fig.~\ref{fig_raa}, are consistent with those of 
\cite{Tribedy:2011aa,Albacete:2012xq}.
The shadowing calculations of Helenius et al. \cite{Helenius:2012wd}, 
performed at NLO with the EPS09s Parton Distribution Functions and DSS 
fragmentation functions describe the data well (the calculations are for 
$\pi^0$).
The predictions by Kang et al. \cite{Kang:2012aa}, performed within a
framework combining leading order (LO) pQCD and cold nuclear matter effects,
show $R_{\mathrm{pPb}}$ values below unity for $p_{\mathrm{T}}\gtrsim 6$ GeV/$c$, 
which is not supported by the data.
The prediction from the HIJING 2.1 model \cite{Xu:2012au} 
describes, with shadowing, the trend seen in the data, although it seems that,
with the present shadowing parameter $s_g$, the model underpredicts 
the data. 
The HIJING model implementation of decoherent hard collisions (DHC) has a small 
influence on the results; the case of independent fragmentation is included 
for this model and improves agreement with data at intermediate $p_{\mathrm{T}}$.
The comparisons in Fig.~\ref{fig_raa} clearly illustrate that the data 
are crucial for the theoretical understanding of cold nuclear matter 
as probed in \mbox{p--Pb} collisions at the LHC. 

In summary, we have reported measurements of the charged-particle 
$p_{\mathrm T}$ spectra and nuclear modification factor in minimum bias (NSD) 
\mbox{p--Pb} collisions at $\sqrt{s_{\mathrm{NN}}} = 5.02$~TeV.  
The data, covering $0.5<p_{\mathrm T}<20$ GeV/$c$, show a  
nuclear modification factor consistent with unity for 
$p_{\mathrm T}\gtrsim 2$ GeV/$c$.
This  measurement indicates that the strong suppression of hadron 
production at high $p_{\mathrm T}$ observed at the LHC in Pb--Pb collisions 
is not due to an initial-state effect, but is the fingerprint of jet 
quenching in hot QCD matter.

%
\newenvironment{acknowledgement}{\relax}{\relax}
\begin{acknowledgement}
\section*{Acknowledgements}

We would like to thank J. Albacete, A. Dumitru, I. Helenius, S. Roesler,
P. Tribedy, R. Venugopalan, I. Vitev, X.-N. Wang, and their collaborators 
for useful input concerning their models.

The ALICE collaboration would like to thank all its engineers and technicians for their invaluable contributions to the construction of the experiment and the CERN accelerator teams for the outstanding performance of the LHC complex.
The ALICE collaboration acknowledges the following funding agencies for their support in building and
running the ALICE detector:
State Committee of Science, Calouste Gulbenkian Foundation from
Lisbon and Swiss Fonds Kidagan, Armenia;
Conselho Nacional de Desenvolvimento Cient\'{\i}fico e Tecnol\'{o}gico (CNPq), Financiadora de Estudos e Projetos (FINEP),
Funda\c{c}\~{a}o de Amparo \`{a} Pesquisa do Estado de S\~{a}o Paulo (FAPESP);
National Natural Science Foundation of China (NSFC), the Chinese Ministry of Education (CMOE)
and the Ministry of Science and Technology of China (MSTC);
Ministry of Education and Youth of the Czech Republic;
Danish Natural Science Research Council, the Carlsberg Foundation and the Danish National Research Foundation;
The European Research Council under the European Community's Seventh Framework Programme;
Helsinki Institute of Physics and the Academy of Finland;
French CNRS-IN2P3, the `Region Pays de Loire', `Region Alsace', `Region Auvergne' and CEA, France;
German BMBF and the Helmholtz Association;
General Secretariat for Research and Technology, Ministry of
Development, Greece;
Hungarian OTKA and National Office for Research and Technology (NKTH);
Department of Atomic Energy and Department of Science and Technology of the Government of India;
Istituto Nazionale di Fisica Nucleare (INFN) and Centro Fermi -
Museo Storico della Fisica e Centro Studi e Ricerche "Enrico
Fermi", Italy;
MEXT Grant-in-Aid for Specially Promoted Research, Ja\-pan;
Joint Institute for Nuclear Research, Dubna;
National Research Foundation of Korea (NRF);
CONACYT, DGAPA, M\'{e}xico, ALFA-EC and the HELEN Program (High-Energy physics Latin-American--European Network);
Stichting voor Fundamenteel Onderzoek der Materie (FOM) and the Nederlandse Organisatie voor Wetenschappelijk Onderzoek (NWO), Netherlands;
Research Council of Norway (NFR);
Polish Ministry of Science and Higher Education;
National Authority for Scientific Research - NASR (Autoritatea Na\c{t}ional\u{a} pentru Cercetare \c{S}tiin\c{t}ific\u{a} - ANCS);
Ministry of Education and Science of Russian Federation,
International Science and Technology Center, Russian Academy of
Sciences, Russian Federal Agency of Atomic Energy, Russian Federal
Agency for Science and Innovations and CERN-INTAS;
Ministry of Education of Slovakia;
Department of Science and Technology, South Africa;
CIEMAT, EELA, Ministerio de Educaci\'{o}n y Ciencia of Spain, Xunta de Galicia (Conseller\'{\i}a de Educaci\'{o}n),
CEA\-DEN, Cubaenerg\'{\i}a, Cuba, and IAEA (International Atomic Energy Agency);
Swedish Research Council (VR) and Knut $\&$ Alice Wallenberg
Foundation (KAW);
Ukraine Ministry of Education and Science;
United Kingdom Science and Technology Facilities Council (STFC);
The United States Department of Energy, the United States National
Science Foundation, the State of Texas, and the State of Ohio.
\end{acknowledgement}

\bibliographystyle{utphys}   %
\bibliography{RpA}
\newpage
\appendix
\section{The ALICE Collaboration}
\label{app:collab}

\begingroup
\small
\begin{flushleft}
B.~Abelev\Irefn{org1234}\And
J.~Adam\Irefn{org1274}\And
D.~Adamov\'{a}\Irefn{org1283}\And
A.M.~Adare\Irefn{org1260}\And
M.M.~Aggarwal\Irefn{org1157}\And
G.~Aglieri~Rinella\Irefn{org1192}\And
M.~Agnello\Irefn{org1313}\And
A.G.~Agocs\Irefn{org1143}\And
A.~Agostinelli\Irefn{org1132}\And
Z.~Ahammed\Irefn{org1225}
% ...
\renewcommand\labelenumi{\textsuperscript{\theenumi}~}
\section*{Affiliation notes}
\renewcommand\theenumi{\roman{enumi}}
\begin{Authlist}
\item \Adef{0}Deceased
\item \Adef{M.V.Lomonosov}Also at: M.V.Lomonosov Moscow State University, D.V.Skobeltsyn Institute of Nuclear Physics, Moscow, Russia
\item \Adef{University of Belgrade, Faculty of Physics and "Vinvca" Institute of Nuclear Sciences, Belgrade, Serbia}Also at: University of Belgrade, Faculty of Physics and "Vinvca" Institute of Nuclear Sciences, Belgrade, Serbia
\end{Authlist}
\section*{Collaboration Institutes}
\renewcommand\theenumi{\arabic{enumi}~}
\begin{Authlist}
\item \Idef{org1332}A. I. Alikhanyan National Science Laboratory (Yerevan Physics Institute) Foundation, Yerevan, Armenia
\item \Idef{org1279}Benem\'{e}rita Universidad Aut\'{o}noma de Puebla, Puebla, Mexico
\item \Idef{org1220}Bogolyubov Institute for Theoretical Physics, Kiev, Ukraine
\item \Idef{org20959}Bose Institute, Department of Physics and Centre for Astroparticle Physics and Space Science (CAPSS), Kolkata, India
\item \Idef{org1262}Budker Institute for Nuclear Physics, Novosibirsk, Russia
\item \Idef{org1292}California Polytechnic State University, San Luis Obispo, California, United States
\item \Idef{org1329}Central China Normal University, Wuhan, China
\item \Idef{org14939}Centre de Calcul de l'IN2P3, Villeurbanne, France
\item \Idef{org1197}Centro de Aplicaciones Tecnol\'{o}gicas y Desarrollo Nuclear (CEADEN), Havana, Cuba
\item \Idef{org1242}Centro de Investigaciones Energ\'{e}ticas Medioambientales y Tecnol\'{o}gicas (CIEMAT), Madrid, Spain
\item \Idef{org1244}Centro de Investigaci\'{o}n y de Estudios Avanzados (CINVESTAV), Mexico City and M\'{e}rida, Mexico
\item \Idef{org1335}Centro Fermi -- Centro Studi e Ricerche e Museo Storico della Fisica ``Enrico Fermi'', Rome, Italy
\item \Idef{org17347}Chicago State University, Chicago, United States
\item \Idef{org1288}Commissariat \`{a} l'Energie Atomique, IRFU, Saclay, France
\item \Idef{org15782}COMSATS Institute of Information Technology (CIIT), Islamabad, Pakistan
\item \Idef{org1294}Departamento de F\'{\i}sica de Part\'{\i}culas and IGFAE, Universidad de Santiago de Compostela, Santiago de Compostela, Spain
\item \Idef{org1106}Department of Physics Aligarh Muslim University, Aligarh, India
\item \Idef{org1121}Department of Physics and Technology, University of Bergen, Bergen, Norway
\item \Idef{org1162}Department of Physics, Ohio State University, Columbus, Ohio, United States
\item \Idef{org1300}Department of Physics, Sejong University, Seoul, South Korea
\item \Idef{org1268}Department of Physics, University of Oslo, Oslo, Norway
\item \Idef{org1312}Dipartimento di Fisica dell'Universit\`{a} and Sezione INFN, Turin, Italy
\item \Idef{org1132}Dipartimento di Fisica dell'Universit\`{a} and Sezione INFN, Bologna, Italy
\item \Idef{org1145}Dipartimento di Fisica dell'Universit\`{a} and Sezione INFN, Cagliari, Italy
\item \Idef{org1315}Dipartimento di Fisica dell'Universit\`{a} and Sezione INFN, Trieste, Italy
\item \Idef{org1285}Dipartimento di Fisica dell'Universit\`{a} `La Sapienza' and Sezione INFN, Rome, Italy
\item \Idef{org1154}Dipartimento di Fisica e Astronomia dell'Universit\`{a} and Sezione INFN, Catania, Italy
\item \Idef{org1270}Dipartimento di Fisica e Astronomia dell'Universit\`{a} and Sezione INFN, Padova, Italy
\item \Idef{org1290}Dipartimento di Fisica `E.R.~Caianiello' dell'Universit\`{a} and Gruppo Collegato INFN, Salerno, Italy
\item \Idef{org1103}Dipartimento di Scienze e Innovazione Tecnologica dell'Universit\`{a} del Piemonte Orientale and Gruppo Collegato INFN, Alessandria, Italy
\item \Idef{org1114}Dipartimento Interateneo di Fisica `M.~Merlin' and Sezione INFN, Bari, Italy
\item \Idef{org1237}Division of Experimental High Energy Physics, University of Lund, Lund, Sweden
\item \Idef{org1192}European Organization for Nuclear Research (CERN), Geneva, Switzerland
\item \Idef{org1227}Fachhochschule K\"{o}ln, K\"{o}ln, Germany
\item \Idef{org1122}Faculty of Engineering, Bergen University College, Bergen, Norway
\item \Idef{org1136}Faculty of Mathematics, Physics and Informatics, Comenius University, Bratislava, Slovakia
\item \Idef{org1274}Faculty of Nuclear Sciences and Physical Engineering, Czech Technical University in Prague, Prague, Czech Republic
\item \Idef{org1229}Faculty of Science, P.J.~\v{S}af\'{a}rik University, Ko\v{s}ice, Slovakia
\item \Idef{org1184}Frankfurt Institute for Advanced Studies, Johann Wolfgang Goethe-Universit\"{a}t Frankfurt, Frankfurt, Germany
\item \Idef{org1215}Gangneung-Wonju National University, Gangneung, South Korea
\item \Idef{org20958}Gauhati University, Department of Physics, Guwahati, India
\item \Idef{org1212}Helsinki Institute of Physics (HIP) and University of Jyv\"{a}skyl\"{a}, Jyv\"{a}skyl\"{a}, Finland
\item \Idef{org1203}Hiroshima University, Hiroshima, Japan
\item \Idef{org1254}Indian Institute of Technology Bombay (IIT), Mumbai, India
\item \Idef{org36378}Indian Institute of Technology Indore, Indore, India (IITI)
\item \Idef{org1266}Institut de Physique Nucl\'{e}aire d'Orsay (IPNO), Universit\'{e} Paris-Sud, CNRS-IN2P3, Orsay, France
\item \Idef{org1277}Institute for High Energy Physics, Protvino, Russia
\item \Idef{org1249}Institute for Nuclear Research, Academy of Sciences, Moscow, Russia
\item \Idef{org1320}Nikhef, National Institute for Subatomic Physics and Institute for Subatomic Physics of Utrecht University, Utrecht, Netherlands
\item \Idef{org1250}Institute for Theoretical and Experimental Physics, Moscow, Russia
\item \Idef{org1230}Institute of Experimental Physics, Slovak Academy of Sciences, Ko\v{s}ice, Slovakia
\item \Idef{org1127}Institute of Physics, Bhubaneswar, India
\item \Idef{org1275}Institute of Physics, Academy of Sciences of the Czech Republic, Prague, Czech Republic
\item \Idef{org1139}Institute of Space Sciences (ISS), Bucharest, Romania
\item \Idef{org27399}Institut f\"{u}r Informatik, Johann Wolfgang Goethe-Universit\"{a}t Frankfurt, Frankfurt, Germany
\item \Idef{org1185}Institut f\"{u}r Kernphysik, Johann Wolfgang Goethe-Universit\"{a}t Frankfurt, Frankfurt, Germany
\item \Idef{org1177}Institut f\"{u}r Kernphysik, Technische Universit\"{a}t Darmstadt, Darmstadt, Germany
\item \Idef{org1256}Institut f\"{u}r Kernphysik, Westf\"{a}lische Wilhelms-Universit\"{a}t M\"{u}nster, M\"{u}nster, Germany
\item \Idef{org1246}Instituto de Ciencias Nucleares, Universidad Nacional Aut\'{o}noma de M\'{e}xico, Mexico City, Mexico
\item \Idef{org1247}Instituto de F\'{\i}sica, Universidad Nacional Aut\'{o}noma de M\'{e}xico, Mexico City, Mexico
\item \Idef{org23333}Institut of Theoretical Physics, University of Wroclaw
\item \Idef{org1308}Institut Pluridisciplinaire Hubert Curien (IPHC), Universit\'{e} de Strasbourg, CNRS-IN2P3, Strasbourg, France
\item \Idef{org1182}Joint Institute for Nuclear Research (JINR), Dubna, Russia
\item \Idef{org1143}KFKI Research Institute for Particle and Nuclear Physics, Hungarian Academy of Sciences, Budapest, Hungary
\item \Idef{org1199}Kirchhoff-Institut f\"{u}r Physik, Ruprecht-Karls-Universit\"{a}t Heidelberg, Heidelberg, Germany
\item \Idef{org20954}Korea Institute of Science and Technology Information, Daejeon, South Korea
\item \Idef{org1160}Laboratoire de Physique Corpusculaire (LPC), Clermont Universit\'{e}, Universit\'{e} Blaise Pascal, CNRS--IN2P3, Clermont-Ferrand, France
\item \Idef{org1194}Laboratoire de Physique Subatomique et de Cosmologie (LPSC), Universit\'{e} Joseph Fourier, CNRS-IN2P3, Institut Polytechnique de Grenoble, Grenoble, France
\item \Idef{org1187}Laboratori Nazionali di Frascati, INFN, Frascati, Italy
\item \Idef{org1232}Laboratori Nazionali di Legnaro, INFN, Legnaro, Italy
\item \Idef{org1125}Lawrence Berkeley National Laboratory, Berkeley, California, United States
\item \Idef{org1234}Lawrence Livermore National Laboratory, Livermore, California, United States
\item \Idef{org1251}Moscow Engineering Physics Institute, Moscow, Russia
\item \Idef{org1322}National Centre for Nuclear Studies, Warsaw, Poland
\item \Idef{org1140}National Institute for Physics and Nuclear Engineering, Bucharest, Romania
\item \Idef{org1165}Niels Bohr Institute, University of Copenhagen, Copenhagen, Denmark
\item \Idef{org1109}Nikhef, National Institute for Subatomic Physics, Amsterdam, Netherlands
\item \Idef{org1283}Nuclear Physics Institute, Academy of Sciences of the Czech Republic, \v{R}e\v{z} u Prahy, Czech Republic
\item \Idef{org1264}Oak Ridge National Laboratory, Oak Ridge, Tennessee, United States
\item \Idef{org1189}Petersburg Nuclear Physics Institute, Gatchina, Russia
\item \Idef{org1170}Physics Department, Creighton University, Omaha, Nebraska, United States
\item \Idef{org1157}Physics Department, Panjab University, Chandigarh, India
\item \Idef{org1112}Physics Department, University of Athens, Athens, Greece
\item \Idef{org1152}Physics Department, University of Cape Town and  iThemba LABS, National Research Foundation, Somerset West, South Africa
\item \Idef{org1209}Physics Department, University of Jammu, Jammu, India
\item \Idef{org1207}Physics Department, University of Rajasthan, Jaipur, India
\item \Idef{org1200}Physikalisches Institut, Ruprecht-Karls-Universit\"{a}t Heidelberg, Heidelberg, Germany
\item \Idef{org1325}Purdue University, West Lafayette, Indiana, United States
\item \Idef{org1281}Pusan National University, Pusan, South Korea
\item \Idef{org1176}Research Division and ExtreMe Matter Institute EMMI, GSI Helmholtzzentrum f\"ur Schwerionenforschung, Darmstadt, Germany
\item \Idef{org1334}Rudjer Bo\v{s}kovi\'{c} Institute, Zagreb, Croatia
\item \Idef{org1298}Russian Federal Nuclear Center (VNIIEF), Sarov, Russia
\item \Idef{org1252}Russian Research Centre Kurchatov Institute, Moscow, Russia
\item \Idef{org1224}Saha Institute of Nuclear Physics, Kolkata, India
\item \Idef{org1130}School of Physics and Astronomy, University of Birmingham, Birmingham, United Kingdom
\item \Idef{org1338}Secci\'{o}n F\'{\i}sica, Departamento de Ciencias, Pontificia Universidad Cat\'{o}lica del Per\'{u}, Lima, Peru
\item \Idef{org1133}Sezione INFN, Bologna, Italy
\item \Idef{org1155}Sezione INFN, Catania, Italy
\item \Idef{org1271}Sezione INFN, Padova, Italy
\item \Idef{org1146}Sezione INFN, Cagliari, Italy
\item \Idef{org1313}Sezione INFN, Turin, Italy
\item \Idef{org1316}Sezione INFN, Trieste, Italy
\item \Idef{org1286}Sezione INFN, Rome, Italy
\item \Idef{org1115}Sezione INFN, Bari, Italy
\item \Idef{org36377}Nuclear Physics Group, STFC Daresbury Laboratory, Daresbury, United Kingdom
\item \Idef{org1258}SUBATECH, Ecole des Mines de Nantes, Universit\'{e} de Nantes, CNRS-IN2P3, Nantes, France
\item \Idef{org1304}Technical University of Split FESB, Split, Croatia
\item \Idef{org1168}The Henryk Niewodniczanski Institute of Nuclear Physics, Polish Academy of Sciences, Cracow, Poland
\item \Idef{org17361}The University of Texas at Austin, Physics Department, Austin, TX, United States
\item \Idef{org1173}Universidad Aut\'{o}noma de Sinaloa, Culiac\'{a}n, Mexico
\item \Idef{org1296}Universidade de S\~{a}o Paulo (USP), S\~{a}o Paulo, Brazil
\item \Idef{org1149}Universidade Estadual de Campinas (UNICAMP), Campinas, Brazil
\item \Idef{org1239}Universit\'{e} de Lyon, Universit\'{e} Lyon 1, CNRS/IN2P3, IPN-Lyon, Villeurbanne, France
\item \Idef{org1205}University of Houston, Houston, Texas, United States
\item \Idef{org20371}University of Technology and Austrian Academy of Sciences, Vienna, Austria
\item \Idef{org1222}University of Tennessee, Knoxville, Tennessee, United States
\item \Idef{org1310}University of Tokyo, Tokyo, Japan
\item \Idef{org1318}University of Tsukuba, Tsukuba, Japan
\item \Idef{org21360}Eberhard Karls Universit\"{a}t T\"{u}bingen, T\"{u}bingen, Germany
\item \Idef{org1225}Variable Energy Cyclotron Centre, Kolkata, India
\item \Idef{org1306}V.~Fock Institute for Physics, St. Petersburg State University, St. Petersburg, Russia
\item \Idef{org1323}Warsaw University of Technology, Warsaw, Poland
\item \Idef{org1179}Wayne State University, Detroit, Michigan, United States
\item \Idef{org1260}Yale University, New Haven, Connecticut, United States
\item \Idef{org15649}Yildiz Technical University, Istanbul, Turkey
\item \Idef{org1301}Yonsei University, Seoul, South Korea
\item \Idef{org1327}Zentrum f\"{u}r Technologietransfer und Telekommunikation (ZTT), Fachhochschule Worms, Worms, Germany
\end{Authlist}
\endgroup

%

\end{document}